# Strange metal from spin fluctuations in a cuprate superconductor

D. J. Campbell[1], M. Frachet[1], V. Oliviero[1], T. Kurosawa[2], N. Momono[3], M. Oda[2], J. Chang[4], D. Vignolles[1], C. Proust[1]*, D. LeBoeuf[1]*

[1] Univ. Grenoble Alpes, INSA Toulouse, Univ. Toulouse Paul Sabatier, EMFL, CNRS, LNCMI, 38000 Grenoble, France

[2] Department of Physics, Hokkaido University, Sapporo 060-0810, Japan

[3] Muroran Institute of Technology, Muroran 050-8585, Japan

[4] Department of Physics, University of Zurich, CH-8057 Zurich, Switzerland

*Corresponding authors. Email: cyril.proust@lncmi.cnrs.fr, david.leboeuf@lncmi.cnrs.fr

**Strange metals challenge our understanding of charge transport in metals. Here, we investigate how the strange metal phase of $La_{2-x}Sr_xCuO_4$ is impacted by a field-induced glassy antiferromagnetic state. Using magnetic fields above 80 T, we discover a strong enhancement of the normal state magnetoresistance when entering the spin glass phase. We demonstrate that the spin glass causes insulating-like upturns in the resistivity inside the pseudogap phase, which resolves the origin of the "metal-insulator" crossover. In addition the strange metal phase appears at low temperatures over an extended range of magnetic fields where magnetic moments fluctuate, and disappears when these moments freeze out. We conclude that the transport properties of the strange metal phase are controlled by magnetic fluctuations persisting at the lowest temperatures.**

While in Fermi liquids, the resistivity follows a $T^2$ temperature dependence, the hallmark of strange metal is a $T$-linear resistivity down to low temperatures. This behavior is now observed in several quantum materials such as unconventional superconductors(*1, 2*), and twisted bilayers of van der Waals materials (*3–5*). In the cuprate superconductor $La_{2-x}Sr_xCuO_4$ (LSCO), a strange metal phase is observed in the overdoped regime (*2*). The slope of the $T$-linear resistivity is controlled by a Planckian relaxation time (*6*), and the magnetoresistance is linear-in-field which follows an anomalous quadrature form (*7*). A general and puzzling feature of the strange metal is a linear-in-$T$ resistivity existing over a wide region of the phase diagram in the limit of $T \rightarrow 0$ (*8*). In contrast, linear resistivity down to $T = 0$ is observed in quantum critical metals, but only at a singular parameter in the phase diagram. Several theoretical models have been proposed to explain this extended range of $T$-linear resistivity, including critical states existing over a wide range of non-thermal tuning parameters (*9, 10*) or coupling of electrons to two-level systems via spatially random interactions (*11*). A common ingredient for theories of strange metals is often the existence of a low energy degree of freedom that can effectively couple to charge carriers down to $T\rightarrow 0$. While the extended range of $T$-linear resistivity in organic and pnictide superconductors arises when a magnetic order is suppressed (*1*), its nature and origin in cuprates superconductors remains elusive (*8*).

To tackle this issue, we employ unprecedentedly high magnetic fields as a novel tuning parameter to explore the physics of strange metals in the cuprate superconductor LSCO. Close to the critical doping of the pseudogap $p^* = 0.19$, we discover that *T*-linear resistivity exists down to T→0 over an extended range of magnetic field which is however disrupted by the slowing down of spin fluctuations. This behavior is best captured in Fig.1A, where we show the experimental false color plot of *n*, the exponent of the in-plane resistivity $\rho = \rho_0 + aT^n$ in the (*T*, *B*) diagram obtained in LSCO at $p = 0.188$. Right after the suppression of superconductivity by a magnetic field, we uncover an extended magnetic field region of linear-in-*T* resistivity down to low temperature. In this region, magnetic fluctuations have been detected by ultrasound (*12*). With increasing field, these magnetic fluctuations gradually slow down, with a diverging magnetic correlation time (defined in Supplementary Materials) as shown in Fig. 1B. This marks the emergence of a glassy antiferromagnetic state referred to as a spin glass. Here we demonstrate that the emergence of the spin glass has two major consequences on the transport properties: it causes the loss of strange metal behavior and it causes resistivity upturns, a phenomenon dubbed "metal-insulator crossover" in LSCO (*13*). The phase diagram of Fig. 1A shows that the strange metal phase of LSCO is tied to the existence of a magnetic fluctuations. Our results emphasize the importance of slow, spatially distributed, local moment fluctuations that typifies glassy magnetism, in the mechanism producing strange metallic behavior in cuprates, as found for instance in the Sachdev-Ye-Kitaev (SYK) model and its extensions (*10, 14, 15*).

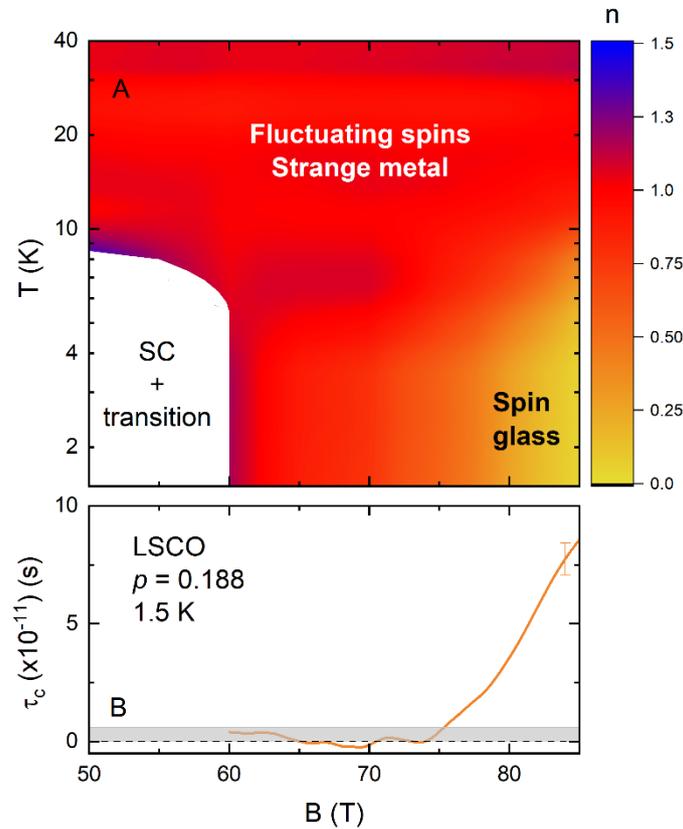

**Fig. 1. Schematic evolution of the resistivity and spin dynamics with temperature and magnetic field in LSCO $p$ = 0.188. (A)** False color plot of the exponent $n$ of the temperature dependent in-plane resistivity $\rho(T) = \rho_0 + aT^n$. It is obtained from interpolation of $(d\ln(\rho(T)-\rho_0)/d\ln T)$ calculated for different magnetic fields. The white area corresponds to the superconducting phase and the resistive transition. $\rho_0$ is the residual resistivity extrapolated to $T=0$ from linear fits as shown in Fig. 3C. The resulting $B$-$T$ diagram shows a $T$-linear regime extending over a wide range of magnetic fields and temperatures, as also shown in the resistivity data of Fig. 3C. However, above $B \approx 70$ T, the $T$-linear regime ends as the system enters the spin glass phase, leading to resistivity upturns. **(B)** Magnetic correlation time, $\tau_c$, as a function of applied magnetic field, calculated from high field ultrasound data measured in the same sample at $T$ = 1.5 K (see Supplementary Materials and fig. S1). As field increases, the magnetic correlation time gradually increases, eventually leading to a freezing of the spins into a spin glass phase. A typical error bar on $\tau_c$ in high field is shown at $B$ = 84 T. The gray shaded area indicates the resolution limit on $\tau_c$. Reducing magnetic field, the fluctuations become faster and at some point, they become too fast to be detected at the ultrasound measurement frequency (typically 100 MHz). Panel (A) and (B) show that the strange metal phase is controlled by spin dynamics and spin freezing suppresses strange metallicity. Magnetic field was applied along the $c$-axis of the crystallographic structure.

**High fields transport measurements in LSCO-** Fig. 2A shows resistivity as a function of magnetic fields up to 86 T at various temperatures in LSCO at $p$ = 0.168. Our data are in quantitative agreement with previous studies performed up to 60 T (*2*, *13*, *16*). At $T$ = 100 K, the magnetoresistance is small and scales with $B^2$. When lowering the temperature, it becomes linear in field (*17*) and can be modeled by a quadrature form (*7*, *18*). The novelty of our study comes from the fact that we extend the measurements to higher magnetic fields and we highlight a novel behavior at low temperatures: the magnetoresistance becomes extremely large, e.g. $\rho(B)$ increases by a factor 3 between $B$ = 60 T and $B$ = 85 T at $T$ = 1.5 K. In this regime, the field dependence of the magnetoresistance is neither linear nor quadratic in $B$, but close to $B^4$ as shown in fig. S2C.

Fig. 2B shows field sweeps of the normalized magnetoresistance at various doping levels from $p$=0.168 to $p$=0.215 at $T$ = 4.2 K. The upturn in the magnetoresistance becomes weaker and appears at higher fields as the doping increases (see also fig. S3 for the raw data). At $p$=0.188 (see fig. S3E and S4C), the magnetoresistance remains linear over an extended field range until an upturn is observed above $B \approx 75$ T. Above $p^*$ ($p$ = 0.215), however, the magnetoresistance remains linear without any upturn up to the highest field of 86 T, in agreement with a previous study (up to 80 T) in this doping regime (*17*).

This is our first main finding: in strong enough magnetic fields, the normal state magnetoresistance of LSCO in the pseudogap phase shows a strong upward deviation from linearity at low temperature.

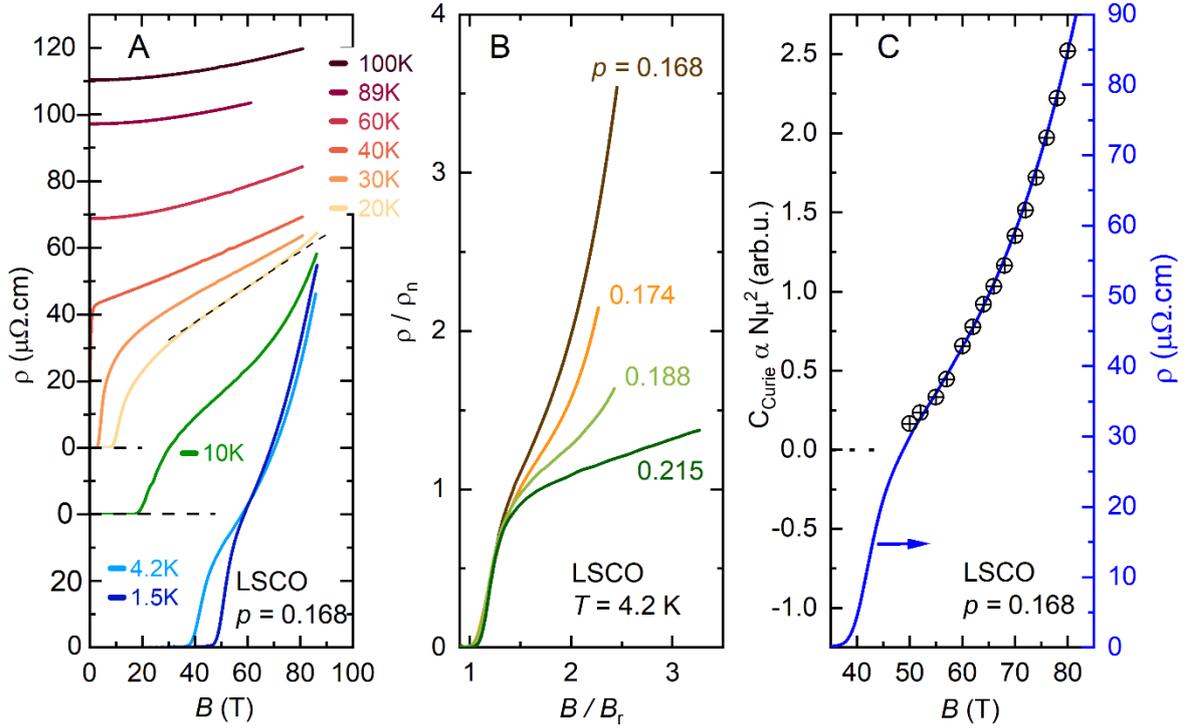

**Fig. 2. Magnetoresistance up to 86 T in LSCO. (A)** In-plane resistivity as a function of magnetic field at different temperatures in LSCO $p$ = 0.168, for field applied along the $c$-axis. Dashed line at 20 K is a linear fit highlighting the upward deviation of the magnetoresistance from linearity for $B$ > 80 T or so. As temperature is decreased further, the magnetoresistance upturn becomes larger and appears at lower fields. **(B)** Evolution of the magnetoresistance at 4.2 K at different doping levels. The resistivity has been normalized at its value right after the resistive transition ($\rho_n$). The resistivity is shown as a function of $B/B_r$ where $B_r$ is the field where the resistivity goes to zero. Raw data are shown in fig. S3 and S4. As doping increases, the magnetoresistance upturn becomes smaller and disappears at $p$ = 0.215 > $p^*$ where it is linear up to the highest measured field. **(C)** Comparison of the magnetoresistance in LSCO $p$ = 0.168 at 4.2 K (blue line) with $C_{Curie}$ (black circles), the Curie constant extracted from fits of the sound velocity data (see Supplementary Materials and fig. S5). In a paramagnet, $C_{Curie}$ is proportional to the number of moments and the size of the moments squared. With increasing field, $C_{Curie}$ increases, as antiferromagnetism develops in the system, in agreement with NMR and neutron data at lower fields (*12, 19, 20*). The scaling between the magnetoresistance and the Curie constant is remarkable and strongly suggests that the magnetoresistance is controlled by the field-dependent magnetism. A similar scaling between $\rho(B)$ and $C_{Curie}$ is obtained in LSCO $p$ = 0.188, see fig. S2B.

**Temperature dependence of the resistivity** – In Fig. 3 we plot $\rho$ vs $T$ at various magnetic fields, obtained from the full raw data set shown in the fig. S3. Close to optimal doping $p$=0.168, a pronounced upturn is observed at low temperature, consistent with prior data at 60 T (*10, 14*). A minimum in the resistivity is clearly observed at $T_{min}$ ~ 25 K at $B$ = 80 T, shifting to lower temperature as the field decreases. This upturn keeps developing up to the highest field, but it is important to note that the system remains metallic with finite resistivity extrapolated at $T \rightarrow 0$. The novelty is that, by measuring up to higher fields, we discovered that the upturn vanishes together with the pseudogap state at $p = p^*$. For $p$ > 0.168, the low-$T$ resistivity upturn gradually weakens and $T_{min}$ shifts to lower temperature. At $p$ = 0.188, close but below $p^*$, there is no insulating-like behavior but only an upward deviation to the $T$-linear resistivity observed below T ≈ 10 K. At $p=0.215 > p^*$, the resistivity remains monotonic without any trace of upturn down to low temperatures and up to the highest fields.

**Origin of the metal-insulator crossover** – In LSCO, when superconductivity is suppressed by a magnetic field, a connection between static magnetism and the pseudogap has been established : the spin glass phase persists well above its zero-field critical doping $p_{sg}$ = 0.135(*21*), actually up to the critical point of the pseudogap phase (*12*). Here we establish a direct connection between this field-induced magnetic freezing and transport properties. First, we rely on the dynamical susceptibility model developed in canonical spin glass system (*22*). Thanks to magneto-elastic coupling, a magnetic susceptibility can be directly deduced from ultrasound measurements: $\chi = \chi_0 + \frac{C_{Curie}}{T}$, where $\chi_0$ is a constant. In a paramagnet, $C_{Curie} \propto N\mu^2$, where $N$ is the number of magnetic moments and μ is the magnetic moment (see Supplementary Materials). This quantity is deduced from fits of sound velocity data (see fig. S5) as previously performed in LSCO $p$ = 0.12 (*23*). In Fig. 2C we compare the magnetoresistance of LSCO $p$ = 0.168 at $T$ = 4.2 K with the field dependence of the parameter $C_{Curie}$. The remarkable scaling from $B$ = 50 T up to the highest fields strongly suggests that the increase in $N\mu^2$ drives the upturn in the magnetoresistance. This scaling is also observed at higher temperature in LSCO $p$ = 0.168 (fig. S2A) and $p$ = 0.188 (fig. S2B). Overall the upturn in the magnetoresistance appears to be governed by the field dependent magnetism, and the underlying picture is that of holes flowing in a fluctuating magnetic environment, which strengthens and slows down as the field increases. Since the sound velocity is proportional to the magnetic susceptibility, the correlation between the magnetoresistance and the sound velocity indicates that the magnetoresistance is driven by the spin susceptibility.

Another connection can be made from a comparison of dρ/dB and the freezing temperature of the spin glass. Fig. 4A-B shows the temperature dependence of dρ/dB at various doping levels, from $p$ = 0.12 up to $p$ = 0.215, above 80 T. In the doping range $p$ < 0.17, a maximum is observed at finite temperature defined as $T_{max}$. Such a behavior is not expected for orbital magnetoresistance which follows the temperature dependence of the mobility. The maximum shifts towards lower temperature at higher doping levels. Note that for $p$ = 0.215 > $p^*$ there is no sign of any enhancement of dρ/dB down to low $T$. In the phase diagram of Fig. 4C, we report the doping dependence of $T_{max}$, which matches remarkably well the freezing temperature of the spin glass, $T_f(p)$, as found by NMR and ultrasound measurements, for all doping levels studied here. Fig. 4C shows that the upturns both in the resistivity and magnetoresistance are tied to the emergence of the spin glass phase. We now understand that the doping level for the metal-insulator crossover is field-dependent because it follows the field-dependence of the spin glass transition. And consequently, since spin freezing ends at $p^*$ when superconductivity is quenched by high fields, the metal-insulator crossover also ends at $p^*$. As demonstrated in Ref. (*12*), if a progressive freezing of magnetic fluctuations would persist at $p$ = 0.215, then it should have been observed at 50 T, well below our maximum field of 86 T. Consequently, spin freezing is a property of the pseudogap and resistivity upturns cannot be observed beyond $p^*$, even at higher fields.

This is our second main finding: the so-called metal-insulator crossover observed in cuprate superconductors originates from the freezing of magnetic fluctuations. Remarkably, Fig. 1 and 2 show that the electronic transport properties are affected by the magnetic field deep inside the non-superconducting state. The upturn in the resistivity is not simply revealed by suppressing superconductivity, but actually induced by magnetic field.

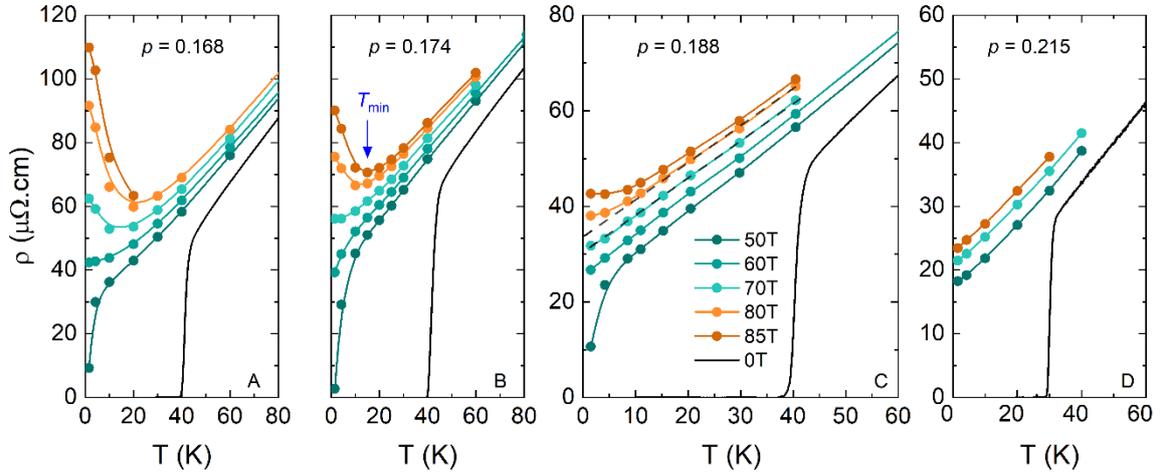

**Fig. 3. Temperature dependence of the in-plane resistivity of LSCO for doping levels across $p^*$ at different magnetic fields. (A-D)** The temperature dependence of the resistivity (circles) is obtained from cuts at fixed fields from the pulsed field data shown in fig. S3. Black line is the resistivity measured at zero field. Colored lines are guides to the eye. Dashed lines in panel C are linear fits to the data. The strong magnetoresistance shown in Fig. 2 produces resistivity upturns at low $T$ for $p < p^*$. As doping increases from $p = 0.168$ to $p = 0.188$, the size of the upturn decreases and the upturn appears also at lower $T$ and higher fields. No upturn is observed in our sample with $p = 0.215$. A marker of this evolution is $T_{min}$ the temperature where $\rho(T)$ has a minimum (blue arrow in panel B in LSCO $p = 0.174$ at 85 T). $T_{min}$ increases with field at a given doping. But as the doping increases, $T_{min}$ becomes lower at a given field. $T_{min}$ is reported in the phase diagram of Fig. 4C. Whereas the resistivity upturn appears as soon as superconductivity is suppressed for $p = 0.168$ and $0.174$, there is an extended range of magnetic field where linear-in-$T$ resistivity is observed at $p = 0.188$. Indeed, between 60 and 70 T the resistivity of LSCO $p = 0.188$ is linear down to the lowest tempratures. At 80 T and above, a clear deviation from linearity is observed at low $T$, as a resistivity upturn starts developing.

**Spin glass phase and superconductivity** – In LSCO the spin glass is stabilized by a magnetic field, in contrast to what is found in canonical spin glass systems, where the magnetic field reduces the freezing temperature and destroys the spin glass phase (*24–26*). Competition between magnetism and superconductivity explains the behavior of LSCO when superconductivity is present (*12*, *19*, *20*, *27*). However, data in high fields indicate that another mechanism is at play. Indeed, within a scenario exclusively based on competition, we would expect magnetic properties to become field-independent above $B_{c2}$. Clearly, this is not what is observed here, as the spin glass phase can be stabilized for fields far above accepted values of $B_{c2}$ (*12*, *28*). This is best seen in fig. S6C where we report the magnetoresistance anomalies along with the onset field for spin freezing determined from ultrasound measurements. Moreover, in LSCO $p = 0.122$, the magnetic properties are still non-saturating at 60 T (*23*), which is three times the estimated value of $B_{c2}$ at this doping (see Supplementary of (*12*)). Finally, the strong enhancement of the magnetoresistance deep within the resistive state cannot be explained by invoking the competition between magnetism and superconductivity alone.

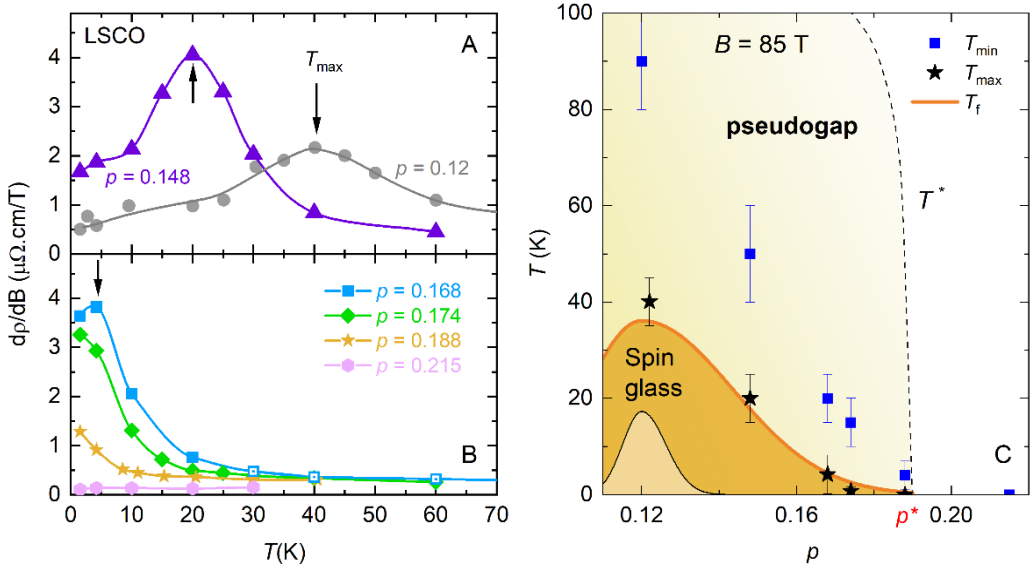

**Fig. 4. Phase diagram of magnetic freezing and resistivity anomalies. (A-B)** Derivative of the magnetoresistance as a function of $T$ measured at 85 T (full symbols) and 80 T (empty squares). d$\rho$/dB is calculated from raw data in fig. S3. Lines are guides to the eye. d$\rho$/dB shows a peak at $T_{max}$ which decreases with increasing doping. For $p$=0.174 and $p$=0.188, there is no maximum in d$\rho$/dB but only an upturn at low $T$. Above $p^*$ the magnetoresistance is much smaller and almost constant up to 30 K as seen in the raw data of fig. S3. **(C)** We report $T_{max}$ (black stars), along with $T_{min}$ (blue squares, defined as the minimum in $\rho(T)$, see Fig. 3) measured at 85 T in the temperature - doping phase diagram. We also plot the freezing temperature, $T_f$, of the spin glass at the ultrasound frequency (orange line) reproduced from Ref.(*12*). The magnetoresistance maximum tracks remarkably well the freezing temperature. We find that $T_{min}$ is proportional to $T_f$. The upturns in $\rho(T)$ (the signature of the metal-insulator crossover (*13*)) and the magnetoresistance are linked to the field-dependent magnetism of LSCO. Note that we also find excellent agreement between the magnetic freezing and the magnetoresistance anomalies in the B-T phase diagram, as shown in fig. S6. Dashed line is the pseudogap temperature $T^*$ from Ref. (*29*). In high magnetic fields, the pseudogap ends at $p^*$ = 0.19 concomitantly with the spin glass and the metal-insulator crossover. The thin black line (light orange area) inside the spin glass dome is a schematic of the boundary of the spin glass phase for $B$ = 0, reproduced from Ref. (*12*). In zero magnetic field, the spin glass phase extends only up to $p_{sg}$=0.135 (*21*). The spin glass phase extends over a wider range of the phase diagram when superconductivity is suppressed (*12, 19, 20*). Since the spin glass phase drives the metal-insulator crossover, the doping where this crossover occurs is field dependent.

**Strange metallicity and magnetic fluctuations** – Having established that the magnetic freezing causes the metal-insulator crossover, we will now discuss how the magnetic fluctuations precursor to freezing are connected to the strange metal phase. It is instructive to look at LSCO at $p = 0.188$ (Fig. 3C). In contrast to lower doping levels where resistivity upturns occur as soon as superconductivity is suppressed (see Fig. 3A and Fig. 3B), we have access to a broader field range where the system is neither superconducting nor magnetically frozen. This situation is rather unique as it occurs only in the close proximity to p*. Between $B$ = 60 T and $B$ = 70 T, the resistivity is linear down to the lowest temperatures. While $\rho_0$, the linear extrapolation of the resistivity at $T \rightarrow 0$, increases significantly with field. the $T$-linear slope only changes slightly. Magnetic field seems to increase elastic scattering. This could be the result of a field-induced increase in disorder level caused by a modification in the local magnetic moments configuration and amplitude (see Fig. 2C and fig. S2B) (*30*). Note that an increase of scattering rate with magnetic field is also observed by spectroscopic probes (*31*). As field keeps increasing, magnetic fluctuations become slower at low temperatures (see Fig. 1B). Above 70 T or so, they presumably become static enough with respect to the timescale of charge carriers (*32*, *33*) and the resistivity deviates from linearity at low $T$. This behavior is best captured in the false color plot of Fig. 1A. It shows that the strange metal behavior is only observed when spins are fluctuating and it is lost when spins are frozen. When spin fluctuations are quenched by a field at low temperatures, the resistivity deviates upwardly from linearity (see Fig. 1A and Fig. 3C).

**Discussion** – NMR and ultrasound studies tell us that magnetic field affects spin dynamics close to $p^*$ (*12*). Here we have shown that this impacts the electronic transport properties in two ways:

First, magnetic field induces freezing of magnetic fluctuations which is at the origin of the metal insulator. Magnetic field usually quenches magnetic fluctuations. This reduces scattering and leads to negative magnetoresistance (*34*). This contrast with the positive magnetoresistance observed here, indicating that another mechanism is at play. The upturn in the magnetoresistance and in the temperature dependence of the resistivity is due to an increase of the scattering rate in the frozen state and / or to a loss of carrier density. The metal-insulator crossover is actually a gradual metal-metal transition. In high fields, this crossover exists up to $p^*$ and is thus tied to the pseudogap phase.

Second, the strange metal phase can be toggled by the magnetic field. We show that the extended domain of criticality, previously observed as a function of doping (*2*), also exists as a function of magnetic field. The essential aspect of magnetism for the strange metal phase is the extended regime of magnetic fluctuations that occur before freezing, not the freezing itself which causes the termination of the strange metal. It is natural to think that the magnetic phase responsible for the extended criticality in the $B$-$T$ diagram is also responsible for the extended criticality in the strange metal phase above $p^*$. While static magnetism disappears at $p^*$, magnetic fluctuations still persist well above $p^*$. Low energy fluctuations are detected beyond $p^*$ with NMR (*35*) and neutron scattering (*36–39*), with low energy spectral weight growing as $T$ is lowered even at $p=0.25$ (*40*). At higher energies, paramagnon excitation are also detected by resonant inelastic x-ray scattering (RIXS) (*41–45*). The magnetic fluctuations found beyond $p^*$ are likely to cause strange metal behavior in this regime, in the same way we found them to be correlated with strange metal behavior at $p = 0.188$ in high fields prior to spin freezing. However, beyond $p^*$ these fluctuations will not become static and form a spin glass. They can consequently sustain strange metallicity up to the highest fields. Indeed, in LSCO $p = 0.215$ and $p = 0.23$ (*2*), a strange metal phase that extends up to the highest field is observed.

The fluctuating magnetic phase underpinning the extended $T$-linear regime as a function of magnetic field is a peculiar magnetic phase. It consists most likely of local moments, with antiferromagnetic exchange interactions, that remain fluctuating even in the T$\rightarrow$0 limit and over an extended range of magnetic field. Such behavior is similar to frustrated magnets where a quantum spin liquid is quenched into a static antiferromagnetic phase with magnetic field(*46*).

The extended fluctuating magnetic regime precursor to the spin glass is reminiscent of a "quantum critical phase", i.e. a critical state spanning a wide range of non-thermal parameter space. Such a phase has been proposed theoretically in order to understand the extended range of $T$-linear behavior of the strange metal(*9*, *47*, *48*). Further investigations are required to understand the exact nature of this fluctuating magnetic phase.

The fluctuating regime precursor to the spin glass is characterized by spatially inhomogeneous local moment fluctuations with increasingly long correlation time (by decreasing $T$ or by increasing $B$). The spatial inhomogeneities observed for instance in the distribution of NMR relaxation time $T_1$ (*12*, *33*) could be the cause for the extended range of criticality as function of doping level or magnetic field, as proposed theoretically (*10*). These peculiar magnetic properties are found, to some extent, in several theoretical models such as the Hubbard model (*30*, *49*, *50*), the SYK-based model (*10*, *14*, *15*) and two-level systems characteristic of spin glasses (*11*, *47*). The later model predicts that an extended strange metal phase can arise when itinerant electrons are interacting with fluctuations of a metallic glass described as a collection of two-level systems. A Fermi liquid state is recovered when the metallic glass is frozen, and this behavior is compatible with the metallic state observed in the spin glass phase of LSCO at low temperatures. However, it remains an open question whether the theory can account for the upturn in resistivity and magnetoresistance. Our work calls for introducing magnetic fields and its impact on spin dynamics in these models in order to understand the magnetotransport properties of strange metals.

**Acknowledgements**. This work was performed at the LNCMI, a member of the European Magnetic Field Laboratory (EMFL). Work at LNCMI was supported by the French Agence National de la Recherche (ANR) Grant No. ANR-19-CE30-0019-01. D.V. and C.P. acknowledges also support from the EUR grant NanoX n° ANR-17-EURE-0009. Work in Zurich is supported by the Swiss National Fundation.

**Author contributions:** T. K., N. M., N. O. grew the LSCO single crystals. D. C., D. L., V. O., D. V., C. P. performed the transport measurements. M. F. performed the fits to the ultrasound data. D. C. and D. L. analysed the experimental data. D. L. and C. P. wrote the manuscript with input from all authors. D. L. supervised the project.

# Supplementary materials

**Materials and Methods**

**Samples** - High quality La$_{2-x}$Sr$_x$CuO$_4$ (LSCO) single crystals were grown by the traveling solvent floating zone method. Samples used in this study are cut from samples previously characterized in high fields with NMR and ultrasound (*1*)

**Determination of hole doping** - The hole doping *p*, which is considered to be equal to the Sr concentration *x* in the absence of oxygen off-stoichiometry, has been determined by measuring T$_{st}$, the temperature of the structural transition from the high-*T* tetragonal (HTT) phase to the low-*T* orthorhombic (LTO) phase by sound velocity.

Since the HTT to LTO transition line goes to 0 at *p* ≈ 0.21, the doping of samples with Sr content above this value is assessed using the thermodynamic bulk superconducting critical temperature *T*$_c$ measured with sound velocity.

**Pulsed field resistivity measurements** - The high field measurements were performed at the pulsed-field facility of the Laboratoire National des Champs Magnétiques Intenses (LNCMI) in Toulouse, France. Measurements up to 86 T were performed in a dual-coil resistive pulsed magnet, each coil being driven by separate and synchronized capacitor banks. Magnetic field was applied along the *c*-axis of the crystallographic structure of LSCO. The in-plane resistivity of LSCO was measured using a conventional four-point configuration with a current excitation between 1 mA and 5 mA at a frequency of ~ 60 kHz. A high-speed acquisition system was used to digitize the reference signal (current) and the voltage drop across the sample at a frequency of 500 kHz. The data were post-analyzed with a software in order to perform the phase comparison. Data for the rise and fall of the field pulse were in very good agreement, thus excluding any heating due to eddy currents. Tests at different frequencies showed excellent reproducibility.

**Modeling of ultrasound data and determination of C$_{Curie}$**

In the 1980's, the ultrasound properties of spin glasses have been thoroughly studied by Doussineau and collaborators. They developed a model to parametrize the ultrasound anomalies around the freezing temperature (*2, 3*). Recently, this model has been shown to successfully reproduce the ultrasound anomalies caused by the spin glass in LSCO *p*= 0.12 (*4*). The parametrization of the ultrasound data in LSCO *p* = 0.12 yields parameters that are in remarkable agreement with what is found by fitting the 1/T$_1$ La NMR relaxation rate(*4*). This, among other facts, demonstrates that ultrasound measurements are sensitive to the magnetic properties of the spin glass phase of LSCO.

The sound velocity is the real part of the complex elastic constant *c*(ω,*T*), which in a spin glass system is expressed as *c*(ω,*T*) = *c*$_0$ [1-*g*$^2$χ(ω,*T*)] where *c*$_0$ is the bare elastic constant, *g* the magneto-elastic coupling and ω the ultrasound frequency. χ(ω,*T*) is the susceptibility given by :

$$\chi(\omega, T) = \int \frac{d\tau}{\tau(T)} \frac{\chi(\omega = 0, T)}{1 + i\omega\tau(T)}$$

with the static susceptibility χ(ω=0,*T*)= χ$_0$ + C$_{Curie}$ / *T* and the magnetic correlation time τ$_c$(*T*) = τ$_\infty$exp(E$_0$/*T*). E$_0$ is an energy scale controlling the thermal activation of the spin fluctuations and τ$_\infty$ is the correlation time for *T* >> E$_0$. τ$_\infty$ is fixed at the value 10$^{-13}$ s as inferred from Ref.(*5*). In order to take into account the inhomogeneity in the spin dynamics that typifies spin glass systems(*2*) (*6*) and which is also found in La-based cuprates(*4*) (*7*) (*8*) (*5*), a Gaussian distribution of $E_0$ is introduced in the model.

Using this model, we can fit the ultrasound data for different magnetic fields and extract both the field dependence of $E_0$ and $g^2C_{Curie}$ (see Fig. S5). Assuming a constant magneto-elastic coupling *g*, we can hence get the field dependence of $C_{Curie}$ as shown in Fig. 2C of the main text and in fig. S2.

**Magnetic correlation time from ultrasound data without fitting the data**

The magnetic correlation time $\tau_c$ discussed above can also be extracted directly from the ultrasound data without making any assumption on its temperature/field dependence. Ignoring the Gaussian distribution of $E_0$, $\tau_c$ can be estimated by combining the sound velocity and ultrasound attenuation data as follows. With the relative change in the complex elastic constant expressed as:

$$\frac{\Delta c}{c} = -g^2 \frac{\chi(\omega = 0)}{1 + i\omega\tau_c}$$

One can extract $\tau_c$ :

$$\tau_c = -\frac{1}{\omega} \Im\left(\frac{\Delta c}{c}\right) \Re\left(\frac{\Delta c}{c}\right)^{-1} \quad Eq. 1$$

The real part of the elastic constant $\Re\left(\frac{\Delta c}{c}\right)$ is linked to the sound velocity, $\frac{\Delta v}{v}$, as follows: $\Re\left(\frac{\Delta c}{c}\right) = 2\frac{\Delta v}{v}$. The imaginary part of the elastic constant $\Im\left(\frac{\Delta c}{c}\right)$ is related to the sound attenuation $\Delta \alpha$ in dB/cm according to this formula: $\Im\left(\frac{\Delta c}{c}\right) = \frac{v}{\omega} \frac{ln(10)}{10} \Delta\alpha$.

In Fig. 1B of the main text, we use this method to estimate $\tau_c(B)$ at 1.5 K in LSCO *p* =0.188. The sound velocity $\frac{\Delta v}{v}(B)$ and attenuation $\Delta\alpha(B)$ used to calculate $\tau_c(B)$ are shown in Fig. S1. Using this method, we neglect the distribution of correlation time arising from the distribution of $E_0$. We can consequently expect quantitative difference with the correlation time extracted from fits containing a distribution of $E_0$. But without a well-defined minimum (maximum) in the sound velocity (attenuation), as observed in LSCO *p* = 0.188, values of $E_0$ (and therefore $\tau_c$) extracted from the fits are unreliable. The simple method exposed here allows us to estimate $\tau_c$ even in absence of a minimum (maximum) in the sound velocity (attenuation) and consequently to get a flavor of the critical slowing down of magnetic fluctuations, as shown in Fig. 1B and Fig. S1C.

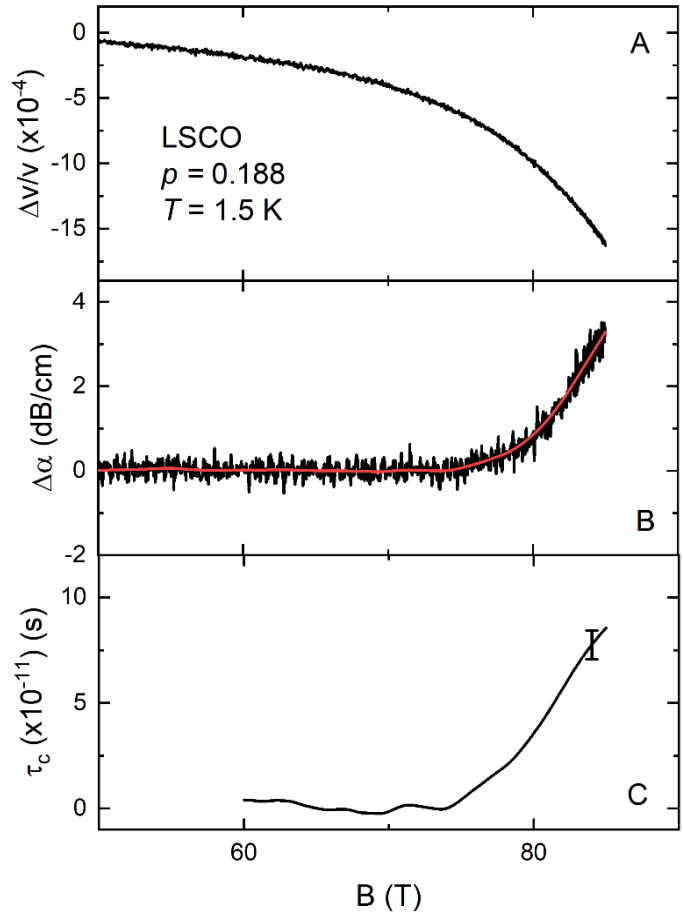

**Fig. S1 – Magnetic correlation time from ultrasound data at 1.5 K in LSCO *p* = 0.188**

**(A)** Sound velocity as a function of magnetic field in LSCO *p* = 0.188 at *T* = 1.5 K. **(B)** Ultrasound attenuation as a function of magnetic field in the same sample at *T* = 1.5 K. Smoothing the data results in the red line. The ultrasound data are reproduced from Ref.(*1*). **(C)** Correlation time $\tau_c(B)$ calculated from sound velocity and attenuation using Eq. 1. It is calculated using the smoothed sound attenuation (red line in panel B). The error bar on $\tau_c(B)$ in high field is ± 7 ps, and mainly comes from noise in the attenuation. Below 60 T or so, the noise on the calculated $\tau_c(B)$ becomes very large, as the calculation involves dividing by the sound velocity which becomes close to 0 as field is reduced.

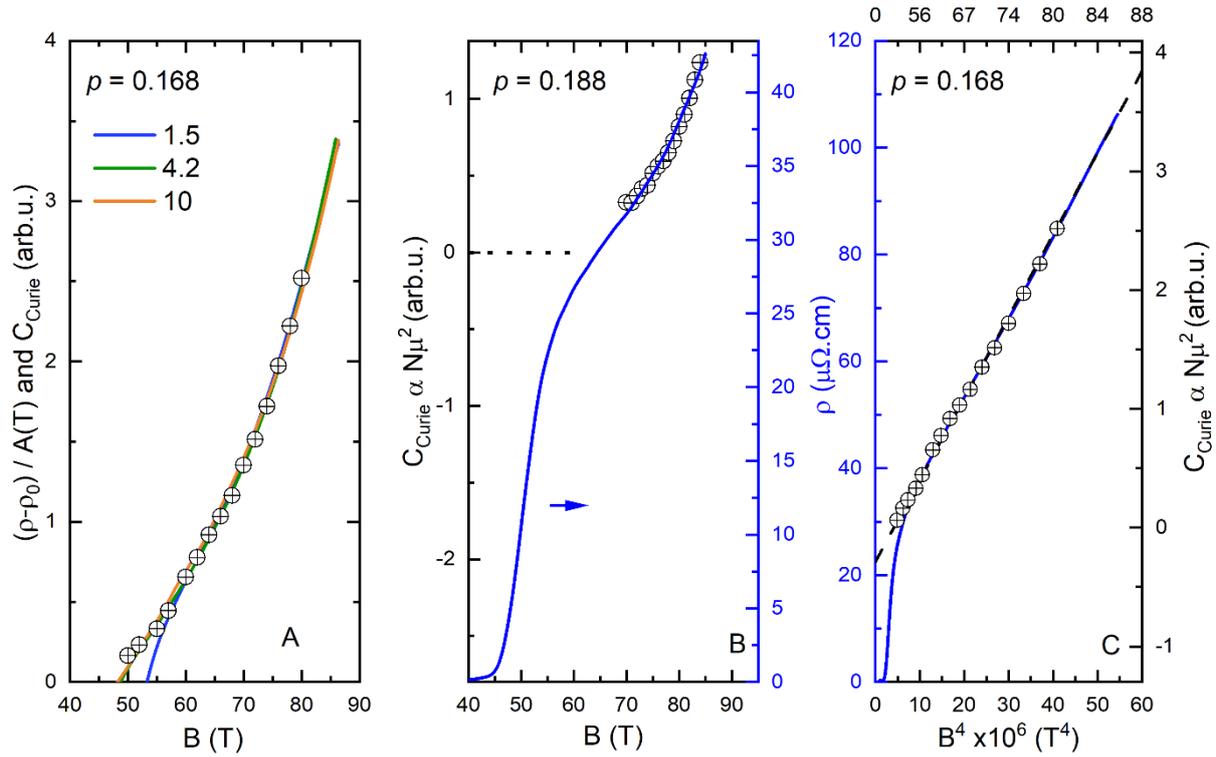

**Fig. S2 – Comparison of magnetoresistance with field-induced magnetism.**

**(A)** Resistivity as a function of magnetic field at different temperatures (colored lines), scaled with Curie constant $C_{Curie}$ (black circles) for doping level $p = 0.168$. The scaling has the form: $\rho = \rho_0 + C_{Curie}(B) \times A(T)$. $A(T)$ is decreasing with increasing $T$. Above 10 K, this simple scaling form does not hold. With increasing $T$, the freezing is shifted to higher $B$, hence the resistivity curves must also be shifted horizontally. **(B)** Resistivity as a function of magnetic field at $T=4.2$ K shown in blue, scaled with Curie constant $C_{Curie}$ (black circles) for doping level and $p = 0.188$ (panel (B)). $C_{Curie}$ is extracted from fits of the sound velocity measured at the same doping levels as explained in the Supplementary Materials text above and in Fig. S5. In both cases, the MR tracks $C_{Curie}$ for fields above the resistive transition. In the paramagnetic state $C_{Curie}$ is proportional to the size of the magnetic moment squared ($\mu^2$) times the number of moments $N$. The MR appears to be controlled by the growth of magnetism with field. **(C)** Same as panel (A) but plotted as a function $B^4$. Both MR and $C_{Curie}$ are linear as a function of $B^4$. Dashed line is a linear fit to the data.

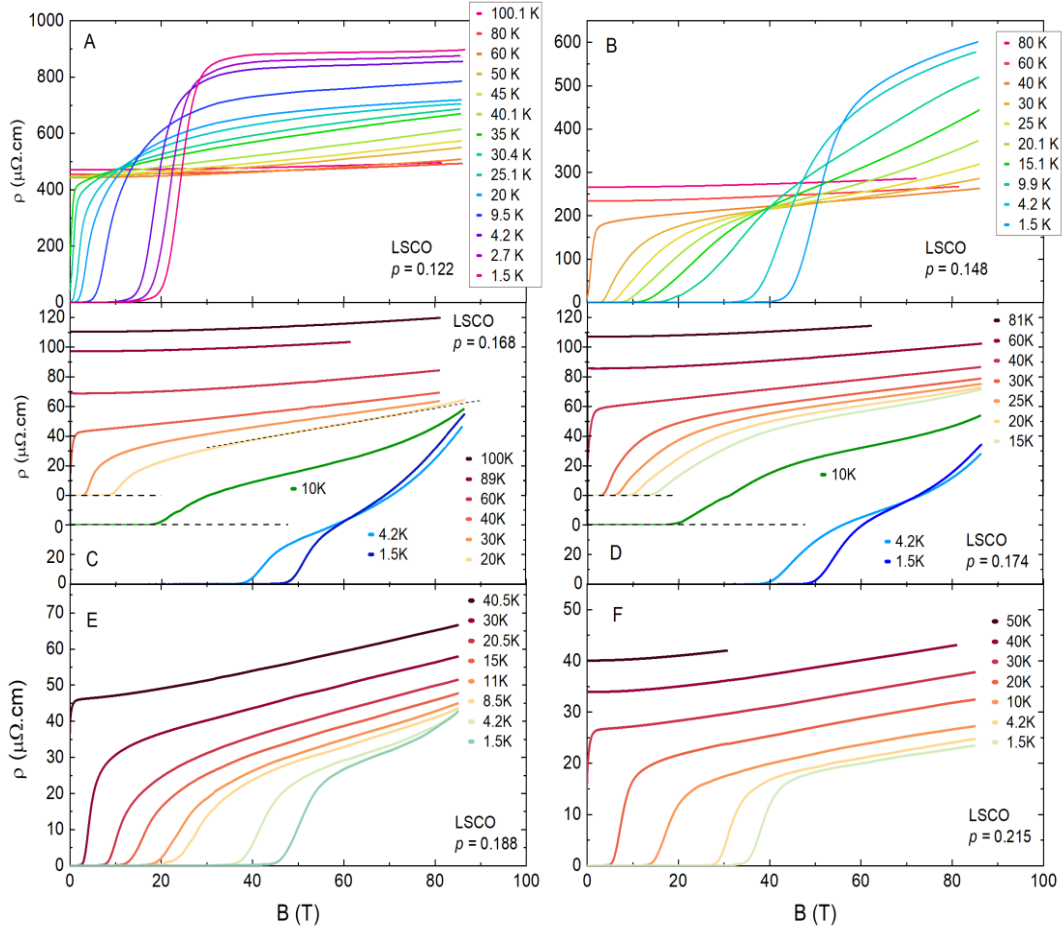

**Fig. S3 – Magnetoresistance ρ(B) for all doping levels studied in this work.**

(A)-(F) In-plane resistivity measured in pulsed field up to 86 T at different temperatures for all doping levels studied in this work. The field is applied along the c-axis. In panel (C) and (D) data below 20 K are shifted vertically for clarity. Cuts of the data at fixed magnetic field yields the temperature dependent resistivity shown in Fig. 3 of the main text. At 40K, the linear magnetoresistance drops continuously with increasing doping level, from 0.34 µΩ.cm/T at $p$ = 0.168, to 0.14 µΩ.cm/T at $p$ = 0.215, in fair agreement with previous reports (*9*) (*10*). Note that the resistivity of our $p$ = 0.122 sample is larger than what can be found in the literature at similar doping level(*11*). This is likely to be the result of c-axis contamination. This however does not impact the qualitative trend of the temperature dependence of the magnetoresistance plotted in Fig. 4A of the main text.

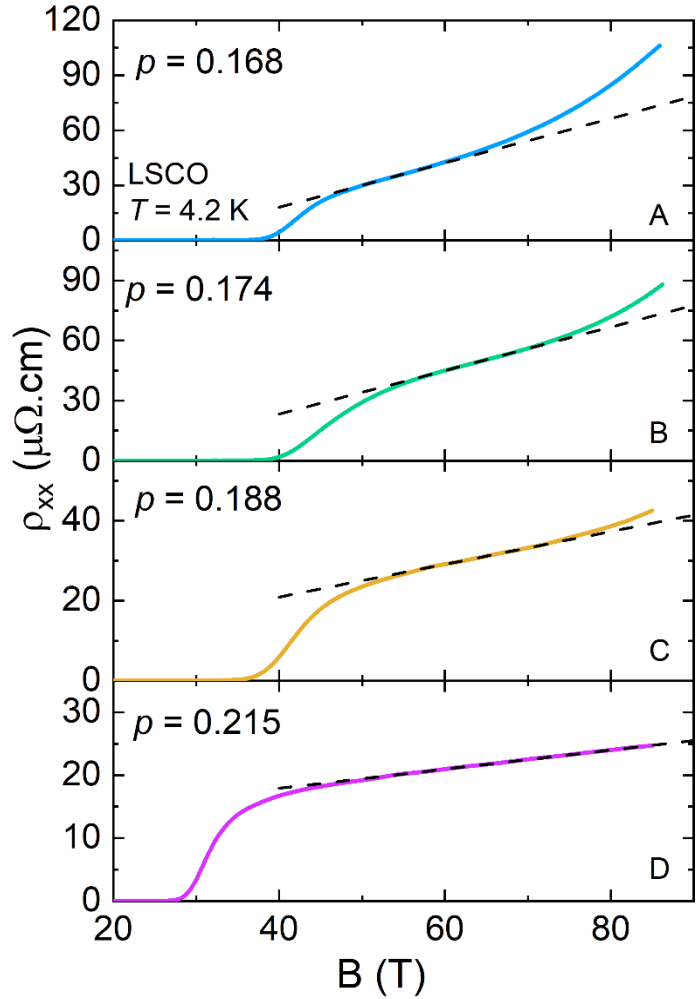

**Fig. S4 – Magnetoresistance of LSCO at various doping levels**

**(A-D)** In-plane resistivity as a function of magnetic field at 4.2 K in LSCO for fields applied along the *c*-axis. Dashed lines are linear fit to the data at intermediate fields made to emphasize on the upward deviation from linearity at high fields. This magnetoresistance upturn appears at higher fields going from *p* =0.168 to *p* = 0.188. Above *p\** = 0.19, no such upturn is observed, in agreement with previous data in 80 T in this doping range (*10*).

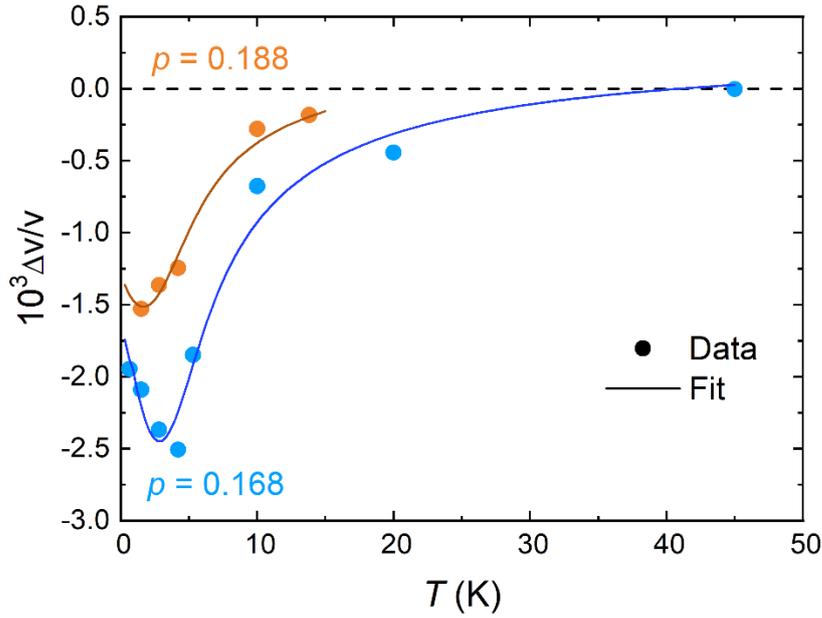

**Fig. S5 – Fits of the ultrasound data for extracting $C_{Curie} \propto N\mu^2$**

Sound velocity data ($\Delta v/v$, circles) of the in-plane transverse mode shown for $p$ = 0.168 (blue, $B$=80 T) and $p$ = 0.188 (orange, $B$=84 T) along with their respective fits (lines). Sound velocity data reproduced from Ref.(*1*). The fits are performed using a phenomenological model described in the text. Fit parameters at $p$ = 0.168 are: $E_0$ = 19.9 K, $\Delta E_0/E_0$ = 0.8, $g^2 \chi_0$ = -8.4 $10^{-4}$, $g^2 C_{Curie}$ = 2.52 $10^{-2}$. Fit parameters at $p$ = 0.188 are: $E_0$ = 13.9 K, $\Delta E_0/E_0$ = 1.2, $g^2 \chi_0$ = -5.4 $10^{-4}$, $g^2 C_{Curie}$ = 1.24 $10^{-2}$. While there is a well-defined minimum in $\Delta v/v$ at $p$=0.168, showing that $E_0$ has a finite value, there is no such minimum in the data at $p$=0.188. Consequently, the data indicate that $E_0$ is zero at $p$=0.188. The value of $E_0$ found by the fitting procedure for $p$=0.188 is thus not reliable. The fit can also be made with fixing $E_0$ = 0 and we can show that this has only a negligible impact on the value $g^2 C_{Curie}$. Importantly it has no impact on the field dependence of $g^2 C_{Curie}$ shown in Fig. S2B.

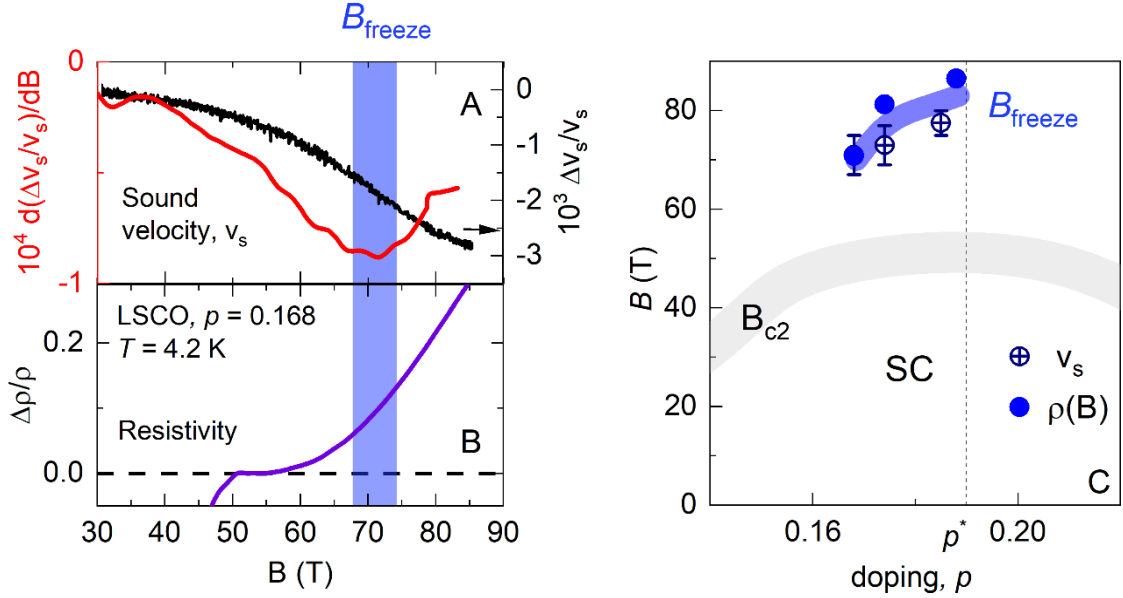

**Fig. S6 – Magnetic field – doping level phase diagram of spin freezing from ultrasound and magnetoresistance**

**(A)** Sound velocity as a function of magnetic field (black, reproduced from Ref. (*1*)) and its field derivative (red). The field-induced slowing down of magnetic fluctuations causes a decrease of the sound velocity of the in-plane transverse mode (*1*). Slowing down is followed by freezing at the ultrasound frequency, as manifested by an inflexion point and a quasi-saturation of the field dependent sound velocity above 70 T or so. Above that field, the temperature dependent sound velocity develops a minimum (see Fig. S5), a signature of freezing in the sound velocity (*1*). We use the inflexion point in Δv/v as a definition for $B_{freeze}$. **(B)** Magnetoresistance plotted as Δρ/ρ, where Δρ is ρ(B) – (aB + ρ$_0$), meaning that we subtracted a linear contribution to ρ(B) corresponding to a linear fit to the raw data in the field range 50 < B < 55 T or so. We see that $B_{freeze}$ (indicated by the blue vertical line) corresponds to a 10% increase in Δρ/ρ. Using that criteria to define the onset field for the magnetoresistance upturn results in the blue solid circles of the B-p phase diagram of panel **(C)**. The MR upturn follows remarkably well $B_{freeze}(p)$ from sound velocity ($v_s$) data, showing that magnetotransport properties are correlated with magnetism. Note that $B_{freeze}$ exceeds the resistively defined $B_{c2}$ (Ref.(*1*) and references therein) in LSCO: the spin dynamics is field-dependent, even when superconductivity is destroyed. Blue line is a guide to the eye.